\documentclass[lettersize,journal]{IEEEtran}
\usepackage{amsmath,amsfonts,amssymb}
\usepackage{algorithmic}
\usepackage{algorithm}
\usepackage{array}
\usepackage[caption=false,font=normalsize,labelfont=sf,textfont=sf]{subfig}
\usepackage{textcomp}
\usepackage{stfloats}
\usepackage{url}
\usepackage{verbatim}
\usepackage{color}
\usepackage{graphicx}
\usepackage{cite}
\begin{document}

\title{Anti-Localization Uplink Communications in Satellite-Terrestrial Systems}

\author{Ranran~Sun, 
Bin~Yang, 
        Yulong~Shen, 
        Yuanyu~Zhang,
       and Xiaohong~Jiang 
\thanks{Ranran Sun is with the Hangzhou Institute of Technology, Xidian University, Hangzhou 311231, China (e-mail: srr\_2013@163.com).}
\thanks{Bin Yang is with the School of Computer and Information Engineering, Chuzhou University, Chuzhou 239000, China (e-mail: yangbinchi@gmail.com).}
\thanks{Yulong Shen is with the School of Computer Science and Technology, Xidian University, Xi'an 710071, China (e-mail: ylshen@mail.xidian.edu.cn).}
\thanks{Yuanyu Zhang is with the School of Computer Science and Technology, Xidian University, Xi'an 710071, China (e-mail: yy90zhang@gmail.com).}
\thanks{Xiaohong Jiang is with the School of Systems Information Science, Future University Hakodate, Hakodate 041-8655, Japan (e-mail: jiang@fun.ac.jp).}}

\markboth{Journal of \LaTeX\ Class Files,~Vol.~14, No.~8, August~2021}%
{Shell \MakeLowercase{\textit{et al.}}: A Sample Article Using IEEEtran.cls for IEEE Journals}


\maketitle

\begin{abstract}
This paper investigates the anti-localization uplink communication in a satellite-terrestrial system, where a ground transmitter Alice communicates with a legitimate satellite receiver Bob in the presence of multiple cooperative adversarial satellites attempting to localize Alice with the time difference of arrival (TDOA) technique. 
Specifically, we propose a cooperative jamming-based scheme for such anti-localization communication, in which Alice exploits the superposition coding with power allocation to simultaneously transmit information/jamming signals for communication with Bob and for confusing signal detection/TDOA measurement at adversarial satellites, while Bob employs the combining vector technique to enhance the desired information signal and also suppress the jamming.
 We define a localization error probability (LEP) metric to jointly depict both the impacts of signal detection and TDOA measurement on localization performance,
and then develop a theoretical framework for the LEP modeling under the proposed scheme.
 We further explore the joint optimal design of jamming coding and power for LEP maximization, subject to the constraints of Alice-Bob communication reliability and Alice's transmit power. 
An effective sample average approximation method is also provided to tackle this non-convex optimization problem.
Finally, extensive numerical results are illustrated to validate our theoretical models and demonstrate how the cooperative jamming helps to provide an anti-localization guarantee while ensuring communication reliability.

\end{abstract}

\begin{IEEEkeywords}
Anti-localization communication, satellite-terrestrial systems, cooperative jamming, time difference of arrival.
\end{IEEEkeywords}

\section{Introduction}
\IEEEPARstart{S}{atellite}-terrestrial communication systems become the indispensable building blocks for future beyond 5G and 6G networks to provide wide-area connectivity guarantee, particularly for remote regions without terrestrial coverage or disaster areas with unreliable or damaged terrestrial infrastructures~\cite{al2022survey,maral2020satellite}. In these systems, the uplink communications from ground terminals to satellites are usually adopted to implement the rapid situation reporting, command-and-control actions and time-sensitive data delivery~\cite{li2024fundamentals}. Thus, such an uplink communication capability is essential for some critical applications like emergency response, public safety missions and disaster relief, where robust bidirectional communications must be ensured even under harsh and highly dynamic network conditions.


As satellite networks continue to evolve toward larger constellations and broader service coverage, existing research efforts on satellite-terrestrial uplink communications are mainly devoted to improving link reliability and/or increasing spectral efficiency through adopting some new techniques such as advanced coding and modulation, adaptive resource allocation, and multi-antenna transmission, etc.~\cite{lin2020secure,huang2023deep,castro2007cross,you2020massive}. However, due to the broadcast nature of wireless propagation, uplink signals from a terrestrial transmitter can also be intercepted by adversarial satellites in the same orbital region of the intended satellite~\cite{heo2023mimo}. By exploiting multiple spatially separated observations, adversary satellites may perform signal detection and apply geolocation techniques such as time difference of arrival (TDOA) positioning to infer the transmitter’s location~\cite{JNaNA1,ho1993solution,kaune2012accuracy,liang2012tdoa}. It is notable that such a location exposure to adversaries may incur consequent persistent tracking, targeted strikes and privacy leakage of the transmitter, posing a significant threat to the security of legitimate users. Thus, location-protected uplink communication against adversarial satellites’ detection and geolocation becomes particularly critical for some location-sensitive applications, such as military, maritime communication, etc.~\cite{bodenhausen2023securing,lorincz2004sensor}.


Conventionally, the location protection of a transmitter is mainly addressed for terrestrial networks at upper application and network layers (see Section II for Related Works), where the techniques of network-layer routing obfuscation and application-layer anonymity/fake-source construction are usually exploited for transmitter location protection. Such upper location protection solutions, however, are not efficient for preventing adversaries from localizing a transmitter at the physical layer, since adversaries can still apply the physical layer sensing and signal-level measurements to infer the transmitter’s location~\cite{torrieri2007statistical,hao2020interference,clements2023dual}. By now, the location protection of a transmitter at the physical layer remains largely unexplored, which motivates us to develop an anti-localization communication framework to protect the transmitter’s location from adversary satellites while maintaining reliable uplink communications in satellite-terrestrial systems.

The main contributions of this paper are summarized as follows.
\begin{itemize}
    \item  We consider a satellite-terrestrial system consisting of a terrestrial transmitter (Alice), a legitimate satellite receiver (Bob), and multiple adversarial satellites (Willies) co-orbiting with Bob, where Willies passively detect the uplink signals from Alice and try to perform TDOA-based geolocation to infer the location of Alice if such signals are detected. To achieve both the reliable Alice-Bob uplink communications and location protection for Alice, we propose a novel cooperative jamming-based anti-localization communication scheme, in which Alice exploits the superposition coding with power allocation to simultaneously transmit information/jamming signals for communication with Bob and for confusing signal detection/TDOA measurement at Willies, while Bob employs the combining vector technique to enhance the desired information signal and suppress the jamming signal at the same time.
    \item  We propose a new metric, namely localization error probability (LEP), to jointly characterize the impacts of signal detection and TDOA measurement on the achievable localization performance at Willies. By applying the hypothesis test for signal detection probability analysis and adopting the generalized cross-correlation and Cramér-Rao lower bound (CRLB) analysis for TDOA estimation error analysis, a comprehensive theoretical framework is then established for LEP modeling under the proposed anti-localization communication scheme. A representative case study is also provided to illustrate the significance of LEP in depicting the localization performance with TDOA.
    \item To degrade as much as possible the localization performance at Willies while ensuring the Alice-Bob communications, we further explore the LEP maximization through the joint optimal design of the maximum transmit power and coding for the cooperative jamming, subject to the constraints of Alice-Bob communication reliability and Alice’s transmit power. An effective algorithm based on the sample average approximation is also provided to tackle this non-convex optimization problem.
    \item Finally, we present extensive numerical results to validate our theoretical models and to demonstrate how cooperative jamming helps to provide an anti-localization guarantee while ensuring communication reliability.
\end{itemize}

The rest of this paper is organized as follows. Section II presents the related works.
Section III introduces the system models and localization performance metric definition.  
Section IV provides theoretical analysis for LEP at Willies. And the LEP maximization is explored in Section V.
We provide the extensive numerical results in Section VI. Finally, Section VII concludes this paper. 

\section{Related Works}
Location protection has been extensively investigated in wireless networks, particularly in wireless sensor networks (WSNs), industrial Internet of Things (IIoT), and programmable wireless environments. 
The methods exploited can be broadly classified into network-layer routing obfuscation, application-layer anonymity/fake-source construction, respectively.

\subsection{Network-Layer Routing Obfuscation}
Routing-based source-location protection aims to prevent adversaries from tracing packet forwarding paths back to the true source. The work~\cite{li2011quantitative} proposed a quantitative framework for measuring and designing source location protection schemes in WSNs, where routing randomness is introduced to increase the uncertainty of the adversary's traceback process. This work provided one of the representative analytical foundations for evaluating source location protection from a routing perspective. 
The work~\cite{mahmoud2011cloud} further investigated source location protection against hotspot-locating attacks and proposed a cloud-based protection scheme in which traffic is distributed through a cloud region to hide the true source location. 
These studies show that routing uncertainty and traffic dispersion are effective for protecting sources against traffic-analysis adversaries. However, they mainly consider network-layer attackers that infer the source by observing packet flows, rather than monitoring nodes that directly detect and localize a transmitter from physical-layer signal measurements.

\subsection{Application-layer Anonymity/Fake-source Construction}
Another line of research exploits anonymity-cloud, multi-sink, and probabilistic forwarding mechanisms to improve source location protection. The work~\cite{wang2019source} proposed an anonymity-cloud-based source location privacy protection scheme, where the source message is divided and disseminated through an irregular anonymity cloud to hide the real source while maintaining data confidentiality and fault tolerance. The work~\cite{han2019cpslp} developed CPSLP, a cloud-based source-location privacy protection scheme using multiple sinks, where packets are forwarded through randomly selected destinations to diversify routing paths and conceal the source location. 
In addition, Wang et al. proposed a probabilistic source-location privacy protection scheme, which randomizes forwarding behavior to balance privacy protection and network overhead~\cite{wang2019probabilistic}. 
For IIoT-oriented WSNs, the work~\cite{han2019dynamic} designed a dynamic multipath source location privacy protection scheme using multiple sinks, further improving path diversity and privacy protection in industrial scenarios.
Although these approaches significantly enhance source-location privacy protection in multi-hop networks, they still rely mainly on modifying routing behaviors or constructing anonymous forwarding regions. Therefore, they are not directly applicable to satellite-terrestrial surveillance systems, where adversarial satellites can detect the signal from Alice and perform TDOA localization without relying on packet routing traces.


\subsection{Localization Performance Analysis}
Recent studies also provide some fundamental methods for the localization performance analysis.
Wymeersch et al. discussed the opportunities and challenges of RIS-enabled radio localization and mapping, emphasizing that programmable environments can reshape propagation paths and improve localization capability~\cite{wymeersch2020radio}. Win et al. further introduced the concept of location awareness via intelligent surfaces and discussed a path toward holographic network localization and navigation~\cite{win2022location}. More recently, RIS-aided localization under hardware impairments, such as pixel failures, has also been studied, showing that localization accuracy is sensitive to both propagation control and system imperfections~\cite{emenonye2023ris}. These studies mainly focus on enhancing localization performance, providing some reference for exploring the localization performance analysis framework.

Existing studies have also investigated TDOA localization performance from the perspectives of estimation accuracy, measurement uncertainty, and system imperfections. \cite{5977569} analyzed passive emitter localization using TDOA measurements in sensor networks and emphasized that the accuracy of TDOA localization is strongly affected by the measurement covariance structure and the sensor-emitter geometry. They derived CRLB expressions for both constant-variance and parameter-dependent-variance models, and used Monte Carlo simulations to compare a static three-sensor network with a moving sensor pair.   
\cite{8964427} considered a more challenging TDOA localization problem where both the signal propagation speed and the true sensor positions are unknown. In their formulation, the measured sensor positions are corrupted by Gaussian position errors, while the TDOA measurements are also modeled with Gaussian noise. They developed an SDP-based algorithm for joint source location and propagation speed estimation. Their performance evaluation used RMSE and CRLB as benchmarks, showing that the proposed SDP approach achieves accuracy close to the CRLB and outperforms a closed-form 2SWLS method, especially when only the minimum number of sensors is available. The study demonstrates that TDOA localization accuracy can be significantly degraded by practical uncertainties such as sensor position errors and unknown propagation speed.

Despite the above progress, existing source location privacy protection and localization-jamming studies cannot directly address the problem considered in this paper. First, most routing-based and anonymity-cloud-based schemes protect against traffic-analysis attacks in terrestrial multi-hop networks, whereas satellite Willies can directly observe Alice's signal and perform physical-layer localization. Second, existing localization-jamming works generally optimize localization degradation under the assumption that localization measurements are always available, while our model considers a hierarchical satellite monitoring protocol in which localization is triggered only after successful detection by the primary Willie. Third, prior works rarely jointly characterize missed detection probability, random auxiliary participation, TDOA estimation error, and final localization error probability under unknown transmitter location.
Motivated by these limitations, this paper develops a cooperative jamming-based approach for transmitter location protection against cooperative satellite-based TDOA localization. 

\section{System Model}
\begin{figure}
	\centering
	\includegraphics[width=3.0in]{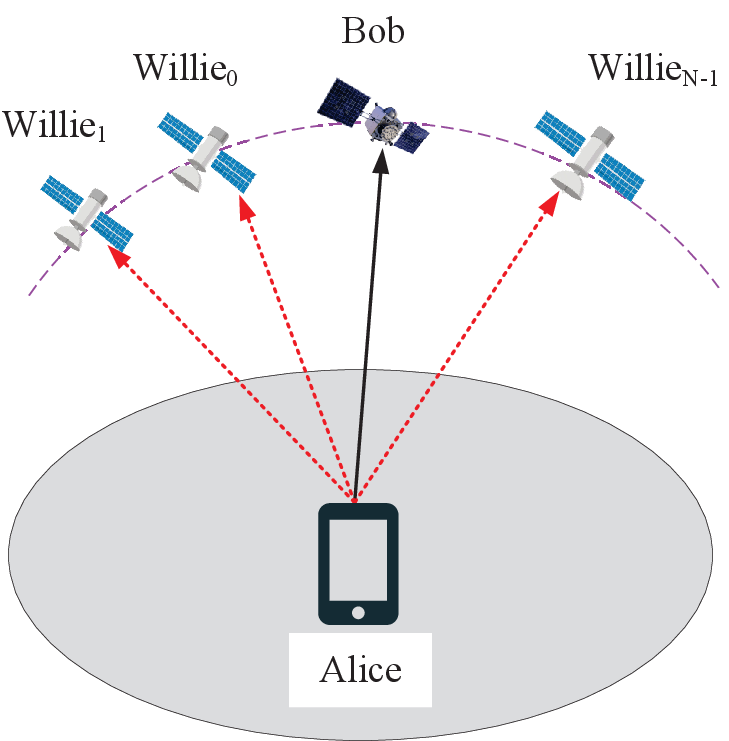}
	\caption{System model.}
	\label{Fig.1}
\end{figure}
As illustrated in Fig.~\ref{Fig.1}, we consider a satellite-terrestrial uplink communication system consisting of a legitimate terrestrial transmitter Alice, an intended satellite receiver Bob, and multiple adversarial satellites $\mathrm{Willie}_i$ that share the same orbit with Bob, where $i\in\{0,1,\ldots,N-1\}$. Alice communicates with Bob, while the adversarial satellites attempt to detect the transmitted signal and further infer the location of Alice based on their observations. 
To protect her position from being localized by Willies, Alice adopts a cooperative jamming-based strategy, where jamming signals are intentionally transmitted along with the communication signal by exploiting the superposition coding with power allocation.
Alice is equipped with $N_A$ antennas, Bob is equipped with $N_B$ antennas, and the $i$-th adversarial satellite is equipped with $N_{W_i}$ antennas.

Among the adversarial satellites, $\mathrm{Willie}_0$ serves as the primary adversarial satellite, while the remaining $N-1$ satellites act as auxiliary adversarial satellites. The primary satellite $\mathrm{Willie}_0$ continuously monitors a suspicious ground region and performs energy detection to determine whether Alice is transmitting. Once $\mathrm{Willie}_0$ declares $\mathcal{H}_1$, indicating that Alice transmits a signal, it forwards the signal fingerprints to the auxiliary satellites through inter-satellite links or ground stations. The auxiliary satellites then tune their receivers according to the received fingerprints and independently perform energy detection. If an auxiliary satellite also declares $\mathcal{H}_1$, its received signal is processed jointly with the signal received at $\mathrm{Willie}_0$ through generalized cross-correlation to estimate the corresponding TDOA. The obtained TDOA measurements are then used to localize Alice. In contrast, if $\mathrm{Willie}_0$ declares $\mathcal{H}_0$, indicating that Alice does not transmit signals, the auxiliary satellites do not perform further detection or localization.

Since the adversarial satellites do not know Alice's true location, Alice is assumed to be located in a suspicious ground region $\mathcal A$, which is approximated by a circular region with radius $R_A$. The potential transmitter locations in this region are modeled as a Poisson point process (PPP). Conditioned on the existence of Alice in $\mathcal A$, Alice can be equivalently regarded as a typical random point in the suspicious region, denoted by $\mathbf q_A\in \mathcal A$. Thus, Alice's two-dimensional ground location $\mathbf {q}_A$ can be expressed as
\begin{equation}
    \mathbf{q}_A=
\begin{bmatrix}
r_A\cos\theta_A,\
r_A\sin\theta_A
\end{bmatrix},
\end{equation}
where $f_{r_A}(r)=\frac{2r}{R_A^2},\quad 0\le r\le R_A$, and $\theta_A\sim U(0,2\pi)$.
The locations of Bob and $\mathrm{Willie}_i$ are denoted by $\mathbf q_B$ and $\mathbf q_i$, respectively. The distances from Alice to Bob and from Alice to $\mathrm{Willie}_i$ are respectively given by $d_B=\Vert\mathbf{q}_A-\mathbf{q}_B\Vert$ and $d_i=\Vert\mathbf{q}_A-\mathbf{q}_i\Vert$.
Since $\mathbf{q}_A$ is unknown to the adversarial satellites, $d_i$ and the instantaneous channel gain are all random from the perspective of Willies. 

\subsection{Transmission at Alice}

Alice simultaneously transmits an information-bearing signal and a jamming signal. Let $s[k]$ denote the normalized information symbol satisfying $\mathbb E[|s[k]|^2]=1$, and let $\mathbf z[k]\in\mathbb C^{d_j\times 1}$ denote the jamming signal vector satisfying $\mathbb{E}[\mathbf{z[k]}\mathbf{z}^{H}[k]]=\mathbf I_{d_j}$, where $d_j$ is the number of jamming streams. The transmitted signal vector of Alice is given by
\begin{equation}
\label{signal:Alice}
    \mathbf x_A[k]=\sqrt{P_s}\mathbf w_s s[k]+\sqrt{P_j}\mathbf W_j \mathbf z[k],
\end{equation}
where $\mathbf x_A[k]\in\mathbb C^{N_A\times 1}$, $P_s$ is the fixed transmit power allocated to the information signal, $P_j$ is the jamming power, $\mathbf w_s\in\mathbb C^{N_A\times 1}$ is the information beamforming vector satisfying $\Vert\mathbf w_s\Vert^2=1$, and $\mathbf W_j\in\mathbb C^{N_A\times d_j}$ is the jamming precoding matrix satisfying $\mathrm{tr}(\mathbf W_j\mathbf W_j^H)=1$.

To increase the uncertainty at the adversarial satellites, Alice adopts a random jamming power which is modeled as a uniformly distributed random variable, i.e., $P_j\sim U(0,P_{j,\max})$, 
where $P_{j,\max}$ denotes the maximum transmit power of jamming signal. Such a random jamming strategy not only decreases Willies' ability to detect Alice's signal, but also degrades the reliability of the TDOA estimation by disturbing the cross-correlation peak.
The instantaneous transmit power at Alice is $\mathbb E[|\mathbf x_A[k]|^2]=P_s+P_j$. Therefore, if Alice is subject to a maximum transmit power $P_A^{\max}$, the peak power constraint is given by $P_s+P_{j,\max}\le P_A^{\max}$.

\subsection{Channel Model}
Due to the dominant line-of-sight (LoS) propagation in satellite-terrestrial links, the Alice-Bob link and the Alice-Willie links are modeled as MIMO Rician fading channels. The channel matrix from Alice to Bob is expressed as $\mathbf H_B=\sqrt{L(d_B)}\widetilde{\mathbf H}_B$, where $\mathbf H_B\in\mathbb C^{N_B\times N_A}$. 
Similarly, the channel matrix from Alice to $\mathrm{Willie}_i$ is given by $\mathbf H_i=\sqrt{L(d_i)}
\widetilde{\mathbf H}_i$, where $\mathbf H_i\in\mathbb C^{N_{W_i}\times N_A}$.
The large-scale path loss is modeled as $L(d)=\beta_0 d^{-\alpha}$, 
where $\beta_0$ is the channel gain at the reference distance, and $\alpha$ is the path-loss exponent.
The small-scale Rician fading matrix $\widetilde{\mathbf H}_i$ is given by 
\begin{equation}
    \widetilde{\mathbf H}_i=
\sqrt{\frac{K_i}{K_i+1}}
\overline{\mathbf H}_i
+
\sqrt{\frac{1}{K_i+1}}
\mathbf G_i,
\end{equation}
where $K_i$ is the Rician factor of the link from Alice to $\mathrm{Willie}_i$, $\overline{\mathbf H}_i$ is the component of LoS channel matrix, and $\mathbf G_i\in\mathbb C^{N_{W_i}\times N_A}$ is the scattered component whose entries independently follow $\mathcal{CN}(0,1)$.
Similarly, the small-scale channel matrix from Alice to Bob is modeled as
\begin{equation}
    \widetilde{\mathbf H}_B=\sqrt{\frac{K_B}{K_B+1}}\overline{\mathbf H}_B+\sqrt{\frac{1}{K_B+1}}\mathbf G_B,
\end{equation}
where entries of $\mathbf G_B\in\mathbb C^{N_B\times N_A}$ independently follow $\mathcal{CN}(0,1)$.
Since Bob and the Willies are located at different positions and observe Alice from different angles, different links may have different Rician factors, i.e., $K_B\neq K_0\neq K_1\neq\cdots\neq K_{N-1}$.

It is worth noting that the information signal and the jamming signal are both transmitted by Alice and propagate through the same physical Alice-$\mathrm{Willie}_i$ channel. Hence, for $\mathrm{Willie}_i$, the information and jamming components share the same channel matrix $\mathbf H_i$, while being mapped to different effective spatial directions due to different transmit beamforming and precoding matrices. Specifically, the effective information channel at $\mathrm{Willie}_i$ is $\mathbf h_{s,i}=\mathbf H_i\mathbf w_s$, 
whereas the effective jamming channel is $\mathbf H_{j,i}=\mathbf H_i\mathbf W_j$.

\subsection{Received Signal at Bob}

The received signal at Bob is given by
\begin{equation}
    \mathbf y_B[k]=\sqrt{P_s}\mathbf H_B\mathbf w_s s[k]+\sqrt{P_j}\mathbf H_B\mathbf W_j\mathbf z[k]+\mathbf n_B[k],
\end{equation}
where $\mathbf n_B[k]\sim\mathcal{CN}(\mathbf 0,\sigma_B^2\mathbf I_{N_B})$ denotes the noise at Bob.

We assume that Bob employs a receive combining vector $\mathbf u_B\in\mathbb C^{N_B\times 1},
\Vert\mathbf u_B\Vert^2=1$, the combined signal is given by $\tilde y_B[k]=\mathbf u_B^H\mathbf y_B[k]$.
Accordingly, the received signal-to-interference-plus-noise ratio (SINR) at Bob is given by
\begin{equation}
\label{SINR_B}
    \gamma_B=\frac{P_s|\mathbf u_B^H\mathbf H_B\mathbf w_s|^2}{P_j\Vert\mathbf u_B^H\mathbf H_B\mathbf W_j\Vert^2
+\sigma_B^2}
\end{equation}
To guarantee the communication quality from Alice to Bob, a minimum SINR requirement should be satisfied, i.e., $\gamma_B\ge \gamma_{\mathrm{th}}$, where $\gamma_{\mathrm{th}}$ is the required SINR threshold at Bob.
\subsection{Received Signal at Willies}
To determine whether Alice transmits or not, an energy detector is employed. Each Willie performs binary a hypothesis test consisting of a null hypothesis $\mathcal{H}_0$ (i.e., Alice does not transmit) and an alternative hypothesis $\mathcal{H}_1$ (i.e., Alice transmits). The received signal at $\text{Willie}_{i}$ under $\mathcal{H}_0$ and $\mathcal{H}_1$ is expressed by
\begin{equation}
\label{received signal at Willie}
    \mathbf y_i[k]=
    \begin{cases}
        \mathbf n_i[k], &\mathcal{H}_0,\\
        \sqrt{P_s}\mathbf H_i\mathbf w_s s[k]+\sqrt{P_j}\mathbf H_i\mathbf W_j\mathbf z[k]+\mathbf n_i[k], &\mathcal{H}_1,
    \end{cases}
\end{equation}
where $\mathbf n_i[k]\sim\mathcal{CN}(\mathbf 0,\sigma_W^2\mathbf I_{N_{W_i}})$ is the receiver noise at $\mathrm{Willie}_i$. We assume that all Willies have the same noise power $\sigma_W^2$.

Each Willie adopts energy detection to determine whether Alice is transmitting. The detection statistic at $\mathrm{Willie}_i$ is defined as $T_i=\frac{1}{M_s}\sum_{k=1}^{M_s}\Vert\mathbf y_i[k]\Vert^2$,
where $M_s$ is the number of received samples used for detection. Given a detection threshold $\lambda_i$, the decision rule at $\mathrm{Willie}_i$ is 
\begin{equation}
T_{i}  \underset{\mathcal{D}_0^{i}}{\stackrel{\mathcal{D}_1^{i}}{\gtrless}} \lambda_i,
\label{detector}
\end{equation}
where $\mathcal{D}_0^{i}$ and $\mathcal{D}_1^{i}$ are the decisions given by $\mathrm{Willie}_i$ supporting $\mathcal{H}_0$ and $\mathcal{H}_1$, respectively. If $T_{i}\geq\lambda_i$, $\mathrm{Willie}_i$ makes the decision $\mathcal{D}_1^{i}$ that Alice transmits signal. Otherwise, the decision $\mathcal{D}_0^{i}$ is made that Alice does not communicate with Bob.
Thus, two types of detection errors may occur. A false alarm occurs when $\mathcal{D}_1^{i}$ is declared while $\mathcal{H}_0$ is true, and the corresponding false alarm probability of $\mathrm{Willie}_i$ is given by $P_{fa,i}=\mathcal{P}\{\mathcal{D}_1^{i}|\mathcal{H}_0\}$. A missed detection occurs when $\mathcal{D}_0^{i}$ is declared while $\mathcal{H}_1$ is true, with missed detection probability $P_{md,i}=\mathcal{P}\{\mathcal{D}_0^{i}|\mathcal{H}_1\}$.

Under $\mathcal{H}_0$, we have 
\begin{equation}
    T_i=\frac{1}{M_s}\sum_{k=1}^{M_s}\Vert\mathbf n_i[k]\Vert^2
\end{equation}
Since $\mathbf n_i[k]\sim\mathcal{CN}(\mathbf 0,\sigma_W^2\mathbf I_{N_{W_i}})$, each normalized noise energy term $\frac{ |n_{i,m}[k]|^2}{\sigma_W^2}$ follows an exponential distribution with unit mean. Therefore, the sum of $M_sN_{W_i}$ independent exponential random variables follows a Gamma distribution~\cite{urkowitz1967energy,atapattu2011energy}. Accordingly, under $\mathcal{H}_0$, we have
\begin{equation}
    \frac{M_s T_i}{\sigma_W^2}=\sum_{k=1}^{M_{s}}\sum_{m=1}^{N_{W_i}}\frac{ |n_{i,m}[k]|^2}{\sigma_W^2}\sim\mathrm{Gamma}(M_sN_{W_i},1).
\end{equation}
where $M_sN_{W_i}$ is the shape parameter and the scale parameter is equal to one.

In this paper, a constant false alarm rate (CFAR) criterion is adopted at each Willie. Specifically, the detection threshold $\lambda_i$ is chosen such that the false alarm probability $P_{fa,i}$ remains fixed at a predefined value $\alpha_i \in (0,1)$, i.e.,
\begin{equation}
P_{fa,i}
= \Pr( T_i \ge \lambda_i \mid \mathcal{H}_0 )= \alpha_i.
\label{eq:PFA}
\end{equation}
Equivalently, we have
\begin{equation}
    \Pr(T_i\le \lambda_i|\mathcal{H}_0)=1-\alpha_i.
\end{equation}

According to the cumulative distribution function of the Gamma distribution, the CFAR-based threshold can be obtained as
\begin{equation}
\lambda_i=\frac{\sigma_W^2}{M_s}\gamma_{\mathrm{reg}}^{-1}\left(1-P_{fa,i},M_sN_{W_i}\right)
\end{equation}
where $\gamma_{\mathrm{reg}}^{-1}(p,a)$ denotes the inverse of the regularized lower incomplete Gamma function with respect to its first argument, i.e.,
\begin{equation}
    \gamma_{\mathrm{reg}}\left(\gamma_{\mathrm{reg}}^{-1}(p,a),a\right)=p.
\end{equation}
Here, the regularized lower incomplete Gamma function is defined as
\begin{equation}
\label{incom_gamma}
    \gamma_{\mathrm{reg}}(x,a)=\frac{1}{\Gamma(a)}\int_0^x t^{a-1}e^{-t}dt,
\end{equation}
where $\Gamma(a)$ is the Gamma function.

Under $\mathcal{H}_1$, the average received energy at $\mathrm{Willie}_i$ is
\begin{equation}
\label{average energy}
    T_{i}=\frac{1}{M_{s}}\sum_{k=1}^{M_{s}}\Vert\sqrt{P_s}\mathbf H_i\mathbf w_s s[k]+\sqrt{P_j}\mathbf H_i\mathbf W_j\mathbf z[k]+\mathbf n_i[k]\Vert^{2}.
\end{equation}
where $\Vert \cdot \Vert_F$ is the Frobenius norm. Then, the missed detection probability at $\mathrm{Willie}_i$ is defined as
\begin{equation}
    P_{md,i}=\Pr(T_i<\lambda_i|\mathcal{H}_1).
\end{equation}
Since Alice's location, the Rician fading channel, and the jamming power are random, we analyze the average missed detection probability, which is obtained by
\begin{equation}
\label{average missed detection}
    \bar {P}_{md,i}=\mathbb E_{\mathbf q_A,\mathbf H_i,P_j}\left[\Pr(T_i<\lambda_i|\mathcal{H}_1)\right].
\end{equation}

If an auxiliary satellite successfully detects Alice's signal, its received signal is processed with that of the primary satellite through generalized cross-correlation. Let the set of auxiliary satellites participating in localization be $\mathcal S\subseteq\{1,2,\ldots,N-1\}$
For any $i\in\mathcal S$, the true TDOA between $\mathrm{Willie}_i$ and $\mathrm{Willie}_0$ is
\begin{equation}
    \tau_{i0}=\frac{\Vert\mathbf q_A-\mathbf q_i\Vert-\Vert\mathbf q_A-\mathbf q_0\Vert}{c},
\end{equation}
where $c$ is the speed of light. Due to the uncertainty caused by receiver noise and random jamming, the estimated TDOA is modeled as
\begin{equation}
\label{estimated TDOA}
    \hat{\tau}_{i0}=\tau_{i0}+\Delta\tau_{i0},
\end{equation}
where $\Delta\tau_{i0}$ denotes the TDOA estimation error. The adversarial satellites then use all TDOA measurements from the participating set $\mathcal S$ to estimate Alice's position via the Chan algorithm, leading to the estimated location of Alice $\hat{\mathbf q}_A=f_{\mathrm{Chan}}\left({\hat{\tau}_{i0}}\right)$.
The corresponding localization error is defined as
\begin{equation}
    e=\Vert\hat{\mathbf q}_A-\mathbf q_A\Vert
\end{equation}

\section{Localization Performance at Willies}

In this section, we first define a performance metric to model the localization performance of adversarial satellites and then analyze the localization performance. 
Different from conventional TDOA localization models where all adversarial nodes are assumed to be available and the measurement geometry is deterministic, the considered adversarial satellites suffer from two coupled uncertainties. First, the auxiliary Willies can participate in localization only when they successfully detect Alice's signal after being triggered by the primary satellite $\mathrm{Willie}_0$. Second, even if a set of auxiliary Willies participates in localization, the TDOA measurement accuracy depends on the unknown Alice location, the corresponding satellite geometry, the Rician fading channels, and the random jamming power. Therefore, the final LEP should be obtained by averaging over all possible participating Willie sets and all underlying random factors. 

\subsection{Performance Metric}
The main objective of this work is to protect Alice's location against cooperative detection and TDOA localization. To this end, we define the localization error probability (LEP) as the primary performance metric.

Under $\mathcal{H}_1$, the adversarial satellites fail to effectively localize Alice in two cases. First, the primary satellite $\mathrm{Willie}_0$ misses Alice's transmission and decides $\mathcal{D}_0^{0}$. Since the auxiliary detection and localization procedure is triggered by $\mathrm{Willie}_0$, this event directly prevents the adversarial satellites from performing TDOA localization. Second, $\mathrm{Willie}_0$ correctly declares $\mathcal{D}_1^{0}$, but the final localization error exceeds a prescribed threshold $d_{th}$ due to auxiliary miss detection, insufficient participating satellites, TDOA estimation errors, or unfavorable localization geometry. Accordingly, the LEP is defined as
\begin{equation}
    P_e=\Pr\left(\mathcal{D}_0^{0}\mid \mathcal{H}_1\right)+\Pr\left(\mathcal{D}_1^{0}, e>d_{th}\mid \mathcal{H}_1\right).
\end{equation}
Since $\Pr\left(\mathcal{D}_0^{0}\mid \mathcal{H}_1\right)=P_{md,0}$ and $\Pr\left(\mathcal{D}_1^{0}\mid H_1\right)=1-P_{md,0}$, 
the LEP can be rewritten as
\begin{equation}
\label{LEP}
    P_e=\mathbb E_{\mathbf q_A,\mathbf H_i,P_j}\left[P_{md,0}+(1-P_{md,0})P_{\mathrm{loc}}\right],
\end{equation}
where $P_{\mathrm{loc}}=\Pr\left(e>d_{th}\mid\mathcal{D}_1^{0},\mathcal{H}_1\right)$
denotes the probability that the localization error exceeds $d_{th}$ after the primary satellite has successfully triggered the localization procedure.

Let the auxiliary Willie set be $\mathcal W_a={1,2,\ldots,N-1}$.
Given that $\mathrm{Willie}_0$ declares $\mathcal{H}_1$, each auxiliary satellite independently performs detection. Let $\mathcal S\subseteq \mathcal W_a$ denote the set of auxiliary satellites participating in TDOA localization. The average detection probability of the $i$-th auxiliary satellite is 
\begin{equation}
    \bar{P}_{d,i}=1-\bar{P}_{md,i}
\end{equation}

Accordingly, under the independent detection assumption, the average probability that the set $\mathcal S$ participates in localization is 
\begin{align}
    \bar{P}_{\mathcal S} &=\mathbb E\left[\prod_{i\in\mathcal S}P_{d,i}(\mathbf q_A,\mathbf H_i,P_j)\prod_{j\in\mathcal W_a\setminus\mathcal S}P_{md,j}(\mathbf q_A,\mathbf H_j,P_j)\right]\notag\\
    &=\prod_{i\in\mathcal S}\bar{P}_{d,i}\prod_{j\in\mathcal W_a\setminus\mathcal S}\bar{P}_{md,j}
\end{align}
where $\bar{P}_{md,j}$ is the average missed detection probability of the $j$-th auxiliary satellite.

By applying the law of total probability, $P_{\mathrm{loc}}$ can be written as
\begin{equation}
    P_{\mathrm{loc}}=\sum_{\mathcal S\subseteq\mathcal W_a}\bar{P}_{\mathcal S}P_{\mathrm{fail}}(\mathcal S),
\end{equation}
where $P_{\mathrm{fail}}(\mathcal S)$ is the localization failure probability associated with the participating set $\mathcal S$ whose localization error exceeds the prescribed threshold $d_{th}$.

For two-dimensional ground localization, at least two independent TDOA measurements are required. Therefore, when $|\mathcal S|<2$, the adversarial satellites are regarded as unable to perform effective localization, and we set $P_{\mathrm{fail}}(\mathcal S)=1$. When $|\mathcal S|\ge 2$, the localization failure probability $P_{\mathrm{fail}}(\mathcal S)$ is determined by whether the localization error exceeds the threshold, i.e., $ P_{\mathrm{fail}}(\mathcal S)=\Pr(e>d_{th})$.
Thus, $P_{\mathrm{loc}}$ can be rewritten as
\begin{equation}
\label{Ploc}
   P_{\mathrm{loc}}=\sum_{\mathcal S\subseteq\mathcal W_a,|\mathcal S|<2}\bar{P}_{\mathcal S}+\sum_{\mathcal S\subseteq\mathcal W_a,|\mathcal S|\ge 2}\bar{P}_{\mathcal S}\Pr(e>d_{th}).
\end{equation}
Substituting~\eqref{Ploc} into~\eqref{LEP} yields

\begin{small}
\begin{align}
    \label{eq:LEP}
    P_e&=\bar{P}_{md,0}+\bar{P}_{d,0}\left[\sum_{\mathcal S\subseteq\mathcal W_a,|\mathcal S|<2}\bar{P}_{\mathcal S}+\sum_{\mathcal S\subseteq\mathcal W_a,|\mathcal S|\ge 2}\bar{P}_{\mathcal S}\Pr(e>d_{th})\right]\notag\\
    &=\bar P_{md,0}+\bar{P}_{d,0}\Bigg[\sum_{\mathcal S\subseteq\mathcal W_a,|\mathcal S|<2}\prod_{i\in\mathcal S}
\bar P_{d,i}\prod_{j\in\mathcal W_a\setminus\mathcal S}\bar P_{md,j}\notag\\
&+\sum_{\mathcal S\subseteq\mathcal W_a,|\mathcal S|\ge 2}
\left(\prod_{i\in\mathcal S}\bar P_{d,i}\prod_{j\in\mathcal W_a\setminus\mathcal S}\bar P_{md,j}\right)\bar P_{\mathrm{fail}}(\mathcal S)\Bigg].
\end{align}
\end{small}where $\bar{P}_{d,0}=1-\bar{P}_{md,0}$ denotes the average detection probability of $\mathrm{Willie}_0$ and $\bar P_{\mathrm{fail}}(\mathcal S)=\mathbb E_{\mathbf q_A,{\mathbf H_i},P_j}\left[\Pr(e>d_{th})\right]$ denotes the average localization failure probability.

The metric LEP $P_e$ captures the detection-localization process of the Willie system. Specifically, $\bar{P}_{md,0}$ characterizes localization failure caused by the miss detection of the primary satellite $\mathrm{Willie}_0$, while $\bar{P}_{d,0}P_{\mathrm{loc}}$ characterizes localization failure after the primary satellite has successfully triggered the localization process. Hence, a larger $P_e$ indicates that Alice achieves stronger location protection against the cooperative Willies. The metric in \eqref{eq:LEP} explicitly captures the joint impact of signal detection performance, cooperative satellite availability, and localization accuracy on the overall location inference capability. As such, it provides a comprehensive and operationally meaningful measure for evaluating localization performance in satellite-terrestrial communication systems.

From the definition of LEP in~\eqref{eq:LEP}, we first derive the expression of average missed detection probability of the $i$-th Willie, and we then can calculate the average probability that $\mathcal{S}$ Willies participate in localization. Finally, the localization performance is measured.

\subsection{Average Missed Detection Probability}

Under $\mathcal{H}_{1}$, according to~\eqref{received signal at Willie}, 
the covariance matrix of $\mathbf y_i[k]$ is given by
\begin{equation}
    \mathbf {Y}_i=P_s\mathbf H_i\mathbf w_s\mathbf w_s^H\mathbf H_i^H+P_j\mathbf H_i\mathbf W_j\mathbf W_j^H\mathbf H_i^H+\sigma_W^2\mathbf I_{N_{W_i}}.
\end{equation}
Accordingly, the mean of energy detection statistic at $\mathrm{Willie}_i$ is given by
\begin{align}
     \mu_i&=\mathbb E[T_i|\mathbf q_A,\mathbf H_i,P_j,H_1]\notag\\
     &=\frac{1}{M_{s}}\sum_{k=1}^{M_{s}}\mathbb {E}[\Vert\mathbf y_i[k]\Vert^{2}]\notag\\
     &=\mathrm{tr}(\mathbf {Y}_i)\notag\\
    &=P_s\Vert\mathbf H_i\mathbf w_s\Vert^2+P_j\Vert\mathbf H_i\mathbf W_j\Vert_F^2+N_{W_i}\sigma_W^2,
\end{align}
The corresponding variance is given by
\begin{equation}
    v_i=\mathrm{Var}(T_i|\mathbf q_A,\mathbf H_i,P_j,\mathcal{H}_{1})=\frac{1}{M_s}\mathrm{tr}(\mathbf {Y}_i^2).
\end{equation}

Since the covariance matrix $\mathbf {Y}_i$ under $\mathcal{H}_{1}$ is generally not a scaled identity matrix, the energy statistic $T_i$ becomes a weighted sum of correlated quadratic terms, whose exact distribution is analytically intractable. For analytical tractability, we approximate the distribution of $T_i$ by an equivalent Gamma distribution using moment matching, which is a commonly adopted approach for approximating weighted sums of chi-square random variables and Gaussian quadratic forms. Specifically, we match the first two moments of $T_i$, yielding $T_i\sim
\mathrm{Gamma}(k_i,\theta_i)$, where $k_i=\frac{\mu_i^2}{v_i}$ and $\theta_i=\frac{v_i}{\mu_i}$~\cite{bodenham2016comparison,satterthwaite1946approximate}. 
Then, the missed detection probability of $\mathrm{Willie}_i$ is given by
\begin{align}
    P_{md,i}(\mathbf q_A,\mathbf H_i,P_j)&=\Pr(T_i<\lambda_i|\mathbf q_A,\mathbf H_i,P_j,\mathcal{H}_{1})\\
    &=\gamma_{\mathrm{reg}}\left(\frac{\lambda_i}{\theta_i},k_i\right),
\end{align}
where $\gamma_{\mathrm{reg}}(\cdot)$ is the regularized lower incomplete Gamma function given by~\eqref{incom_gamma}.
 Hence, the average missed detection probability of $\mathrm{Willie}_i$ is defined as
 \begin{align}
 \label{eq:Pmd}
     \bar P_{md,i}=\mathbb E_{\mathbf q_A,\mathbf H_i,P_j}\left[\gamma_{\mathrm{reg}}\left(\frac{\lambda_i}{\theta_i},k_i\right)\right].
 \end{align}
The corresponding average detection probability is $\bar P_{d,i}=1-\bar P_{md,i}$. 
\subsection{TDOA Measurement}
In practice, the TDOA is estimated through generalized cross-correlation. Let $\bar y_i(t)$ and $\bar y_0(t)$ denote the processed received signals at $\mathrm{Willie}_i$ and $\mathrm{Willie}_0$, respectively. The generalized cross-correlation function is given by
\begin{equation}
    R_{i0}(\tau)=\int_{-\infty}^{+\infty}\Psi_{i0}(f)\bar Y_i(f)\bar Y_0(f)e^{j2\pi f\tau}df,
\end{equation}
where $\bar Y_i(f)$ and $\bar Y_0(f)$ are the Fourier transforms of $\bar y_i(t)$ and $\bar y_0(t)$, respectively, and $\Psi_{i0}(f)$ is the weighting function. The TDOA estimate is obtained from the peak location of the cross-correlation function
\begin{equation}
\label{TDOA}
    \hat{\tau}_{i0}=\arg\max_{\tau} R_{i0}(\tau).
\end{equation}

As the modeling of estimated TDOA $\hat{\tau}_{i0}$ in~\eqref{estimated TDOA}, 
under non-ideal but sufficiently regular conditions, $\Delta\tau_{i0}$ can be modeled as a zero-mean Gaussian random variable
\begin{equation}
    \Delta\tau_{i0}\sim\mathcal N(0,\sigma_{\tau,i0}^2).
\end{equation}
However, the variance $\sigma_{\tau,i0}^2$ is not fixed in the considered system~\cite{8964427}. Instead, it depends on Alice's random location, the corresponding Rician channels, and the random jamming power.
According to the CRLB of time-delay estimation, the variance $\sigma_{\tau,i0}^2$ can be modeled as 
\begin{equation}
\label{variance}
    \sigma_{\tau,i0}^2=\frac{\kappa}{B^2\gamma_{i0}},
\end{equation}
where $B$ denotes the effective signal bandwidth, $\kappa$ is a constant that depends on the waveform and estimator, and $\gamma_{i0}$ is the effective SINR for the TDOA measurement. Since TDOA estimation is performed through the correlation between the received signals at $\mathrm{Willie}_i$ and $\mathrm{Willie}_0$, the effective SINR for the TDOA measurement can be modeled as
\begin{equation}
\label{eq_tem}
    \gamma_{i0}=\left(\frac{1}{\gamma_i}+\frac{1}{\gamma_0}\right)^{-1},
\end{equation}
where $\gamma_i$ and $\gamma_0$ are the SINR at the $i$-th Willie and the primary Willie, they are given by
\begin{equation}
    \gamma_i=\frac{P_s\Vert\mathbf H_i\mathbf w_s\Vert^2}{P_j\Vert\mathbf H_i\mathbf W_j\Vert_F^2+N_{W_i}\sigma_W^2},
\end{equation}
and
\begin{equation}
    \gamma_0=\frac{P_s\Vert\mathbf H_0\mathbf w_s\Vert^2}{P_j\Vert\mathbf H_0\mathbf W_j\Vert_F^2+N_{W_0}\sigma_W^2},
\end{equation}
respectively.
Therefore, substituting $\gamma_{i}$ and $\gamma_{0}$ into~\eqref{eq_tem}, we have

\begin{small}
\begin{equation}
    \sigma_{\tau,i0}^2=\frac{\kappa}{B^2}\Bigg(\frac{P_j\Vert\mathbf H_i\mathbf W_j\Vert_F^2+N_{W_i}\sigma_W^2}{P_s\Vert\mathbf H_i\mathbf w_s\Vert^2}+\frac{P_j\Vert\mathbf H_0\mathbf W_j\Vert_F^2+N_{W_0}\sigma_W^2}{P_s\Vert\mathbf H_0\mathbf w_s\Vert^2}\Bigg).
\end{equation}
\end{small}

This indicates that Alice's jamming signal increases the TDOA error variance by reducing the effective SINR at both the auxiliary Willie and the primary Willie. Moreover, since $\mathbf{H}_{i}$ and $\mathbf{H}_{0} $ depend on Alice’s unknown location, the TDOA error variance is also random from the perspective of the Willies. 
Define the range difference measurement error as
\begin{equation}
    \Delta r_{i0}=c\Delta\tau_{i0}.
\end{equation}
 Then
 \begin{equation}
 \label{eq:var}
     \Delta r_{i0}\sim \mathcal N(0,\sigma_{r,i0}^2),
 \end{equation}
 where $\sigma_{r,i0}^2=c^2\sigma_{\tau,i0}^2$.

\subsection{Localization Error Probability}
For the participating set $\mathcal S={i_1,i_2,\ldots,i_M}$, after the TDOA is obtained according to~\eqref{TDOA}, Willies have to solve‌ the following simultaneous equations to determine the location of Alice. 
\begin{equation}
\label{TDOA equations}
    \Vert\mathbf{q}_{A}-\mathbf{q}_{i}\Vert-\Vert\mathbf{q}_{A}-\mathbf{q}_{0}\Vert=c\hat{\tau}_{i0}, i\neq 0 \quad \mathrm{and} \quad i\in \mathcal{S}.
\end{equation}
Due to the nonlinear nature of the TDOA equations~\eqref{TDOA equations}, directly solving them to obtain Alice's location is generally difficult and may suffer from high computational complexity~\cite{kaune2011accuracy}. 
To address this issue, the Chan algorithm is adopted to estimate Alice's position from the available TDOA measurements.
As a representative closed-form estimator for hyperbolic localization, the Chan algorithm has been widely adopted due to its low computational complexity and favorable localization accuracy in practical TDOA systems~\cite{chan1994simple}.

In the Chan algorithm, the covariance of localization error $\hat{\mathbf{q}}_{A}-\mathbf{q}_{A}$ is formulated as
\begin{equation}
    \mathbf C_{\mathcal S}=\left(\mathbf G_{\mathcal S}^T\mathbf R_{\mathcal S}^{-1}\mathbf G_{\mathcal S}\right)^{-1},
\end{equation}
where $\mathbf G_{\mathcal S}$ is the Jacobian matrix, its element is
the gradient of function $r_{i0}(\mathbf q_{A})=\Vert\mathbf{q}_{A}-\mathbf{q}_{i}\Vert-\Vert\mathbf{q}_{A}-\mathbf{q}_{0}\Vert$ in terms of $\mathbf{q}_{A}$, i.e., $\mathbf G_{\mathcal S}=\left[\mathbf g_{i_1 0}^T, \mathbf g_{i_2 0}^T, \vdots, \mathbf g_{i_M 0}^T\right]^{T}$, where $\mathbf g_{i_1 0}^T$ is given by
\begin{equation}
    \mathbf g_{i0}^T=\left.\frac{\partial r_{i0}(\mathbf q)}{\partial \mathbf q}\right|_{\mathbf q=\mathbf q_A}=\frac{(\mathbf q_A-\mathbf q_i)^T}{\Vert\mathbf q_A-\mathbf q_i\Vert}-\frac{(\mathbf q_A-\mathbf q_0)^T}{\Vert\mathbf q_A-\mathbf q_0\Vert}.
\end{equation}
$\mathbf R_{\mathcal S}$ is the covariance matrix of the range difference measurement error $r_{i0}(\mathbf q_{A})$, which is given by
\begin{equation}
    \mathbf R_{\mathcal S}=\mathrm{diag}\left(\sigma_{r,i_1 0}^2,\sigma_{r,i_2 0}^2,\ldots,\sigma_{r,i_M 0}^2\right),
\end{equation}
where $\sigma_{r,i0}^{2}$ is given in~\eqref{eq:var}.

According to the definition of location error probability in~\eqref{eq:LEP}, to measure this metric, we need to derive the average localization failure probability $\bar{P}_{fail}(\mathcal{S})$, which is defined as $\bar P_{\mathrm{fail}}(\mathcal S)=\mathbb E_{\mathbf q_A,{\mathbf H_i},P_j}\left[\Pr(e>d_{th})\right]$. 

Generally, the localization error $\mathbf{e}=\hat{\mathbf{q}}_{A}-\mathbf{q}_{A}$ is approximated as an isotropic Gaussian vector, i.e., $\mathbf{e}\sim\mathcal N(\mathbf 0,\sigma_{\mathcal S}^2\mathbf I_2)$, where $\sigma_{\mathcal S}^2=\frac{1}{2}\mathrm{tr}\left(\mathbf C_{\mathcal S}\right)$. Then $e$ follows a Rayleigh distribution. Therefore, the average localization failure probability is given by
\begin{align}
\label{eq:fail}
    \bar{P}_{\mathrm{fail}}(\mathcal S)&=\mathbb {E}_{\mathbf q_A,{\mathbf H_i},P_j}\left[P_{\mathrm{fail}}(\mathcal S)\right]\notag\\&=\mathbb {E}_{\mathbf q_A,{\mathbf H_i},P_j}[\Pr(e>d_{th})]\notag\\&=\mathbb {E}_{\mathbf q_A,{\mathbf H_i},P_j}\left[\exp\left(-\frac{d_{th}^2}{\mathrm{tr}(\mathbf C_{\mathcal S})}\right)\right],\, |\mathcal S|\ge 2.
\end{align}
When $|\mathcal S|<2$, the participating satellites do not have enough independent TDOA measurements to perform two-dimensional localization. Hence,
\begin{equation}
    \bar P_{\mathrm{fail}}(\mathcal S)=1,\quad |\mathcal S|<2.
\end{equation}

By substituting~\eqref{eq:Pmd} and~\eqref{eq:fail} into~\eqref{eq:LEP}, we can obtain the LEP. Due to the complexity of~\eqref{eq:Pmd} and~\eqref{eq:fail}, the closed-form expression of localization error probability cannot be determined. We will illustrate the numerical results in the Section VI to validate the correctness of the theoretical analysis.

\subsection{Case Study}
To obtain further insight, we consider a special case where all auxiliary Willies have the same average detection probability, i.e., $\bar P_{d,i}=\bar P_d,\, i=1,\ldots,N-1$. 
Then, the number of participating auxiliary Willies $M=|\mathcal S|$ approximately follows a binomial distribution, i.e., $\Pr(M=m)=\binom{N-1}{m}\bar P_d^m(1-\bar P_d)^{N-1-m}$. 

If all participating sets with $m$ Willies participating localization have approximately the same average localization failure probability, denoted by $\bar P_{\mathrm{fail}}(m)$, then
\begin{align}
     \bar P_{\mathrm{loc}}&=
\sum_{m=0}^{1}
\binom{N-1}{m}
\bar P_d^m(1-\bar P_d)^{N-1-m}
\notag\\&+
\sum_{m=2}^{N-1}
\binom{N-1}{m}
\bar P_d^m(1-\bar P_d)^{N-1-m}
\bar P_{\mathrm{fail}}(m).
\end{align}
Accordingly,
\begin{align}
P_{e}
&=\bar P_{md,0}+(1-\bar{P}_{md,0})\Bigg[
\sum_{m=0}^{1}
\binom{N-1}{m}
\bar P_d^m(1-\bar P_d)^{N-1-m}
\notag\\&+
\sum_{m=2}^{N-1}
\binom{N-1}{m}
\bar P_d^m(1-\bar P_d)^{N-1-m}
\bar P_{\mathrm{fail}}(m)
\Bigg].
\end{align} 
This simplified expression clearly shows that stronger jamming reduces $\bar P_d$, which shifts the distribution of the participating auxiliary number $M$ toward smaller values. Meanwhile, stronger jamming also increases the TDOA error variance and hence enlarges $\bar P_{\mathrm{fail}}(m)$. These two effects jointly increase the LEP $P_e$.

From this special case we can have following insights into the impact of random jamming on the localization capability of  Willies.
First, the number of participating auxiliary Willies follows a binomial distribution, whose success probability is determined by the average detection probability $\bar P_d$. As the jamming power increases, $\bar P_d$ decreases, and thus the probability of $M=m$ shifts toward smaller values. Consequently, the probability that the number of participating satellites has fewer than two auxiliary satellites increases, which directly leads to localization failure in two-dimensional TDOA localization. This indicates that Alice's jamming signal can protect her location not only by degrading measurement accuracy, but also by preventing the formation of a sufficient localization geometry.

Second, even when the number of participating auxiliary Willies is sufficient, i.e., $M\ge 2$, the localization failure probability still increases with the jamming power. This is because random jamming reduces the effective SINR at both the primary Willie and the auxiliary Willies, which enlarges the TDOA estimation error variance and hence increases $\mathrm{tr}(\mathbf C_m)$. Since the localization failure probability contains the term $\exp\left(-\frac{d_{th}^2}{\mathrm{tr}(\mathbf C_m)}\right)$, 
a larger $\mathrm{tr}(\mathbf C_m)$ results in a larger probability that the localization error exceeds the threshold $d_{th}$. Therefore, jamming degrades the localization performance even when the Willie system can collect enough TDOA measurements.

 \section{Localization Error Probability Maximization}

This section investigates the LEP maximization problem to determine the optimal cooperative jamming design at Alice. In this work, we focus on protecting Alice's location against cooperative localization attacks while ensuring that the Alice-Bob communication link satisfies the required quality-of-service constraint. 
According to the SINR given in~\eqref{SINR_B}, since $P_j\sim U(0,P_{j,\max})$, the worst-case SINR occurs at $P_j=P_{j,\max}$. Therefore, a robust communication constraint can be imposed as
\begin{equation}
   \gamma_B(P_{j,\max},\mathbf W_j)\ge\gamma_{\mathrm{th}}, 
\end{equation}
where $\gamma_{\mathrm{th}}$ is the minimum SINR requirement at Bob. Equivalently,
\begin{equation}
    P_{j,\max}\Vert\mathbf u_B^H\mathbf H_B\mathbf W_j\Vert^2\le\frac{P_s|\mathbf u_B^H\mathbf H_B\mathbf w_s|^2
}{\gamma_{\mathrm{th}}}-\sigma_B^2.
\end{equation}
In addition, Alice's peak transmit power constraint is $P_s+P_{j,\max}\le P_A^{\max}$.
The jamming precoder is normalized as $\mathrm{tr}(\mathbf W_j\mathbf W_j^H)=1$.

Thus, the LEP maximization problem can be formulated as
\begin{subequations}
  \label{P1}
\begin{align}
\mathbf P_1:\quad
\max_{P_{j,\max},\mathbf W_j}\quad
&P_e(P_{j,\max},\mathbf W_j)\label{P1:obj}\\
\mathrm{s.t.}\quad
&\gamma_B(P_{j,\max},\mathbf W_j)\ge \gamma_{\mathrm{th}},\\
&P_s+P_{j,\max}\le P_A^{\max},\\
&P_{j,\max}\ge 0,\\
&\mathrm{tr}(\mathbf W_j\mathbf W_j^H)=1.
\end{align}
\end{subequations}

Due to the complexity and intractability of the objective function~\eqref{P1:obj}, $\mathbf P_1$ is non-convex. 
Since Alice's exact location is unknown to the Willies, the objective $P_e$ should be evaluated statistically. To obtain a tractable optimization problem, we adopt the sample average approximation (SAA). Specifically, we generate $L$ independent samples
\begin{equation}
\left\{
\mathbf q_A^{(\ell)},
\mathbf H_B^{(\ell)},
\mathbf H_i^{(\ell)},
P_j^{(\ell)}
\right\}_{\ell=1}^{L},
\end{equation}
where $\mathbf q_A^{(\ell)}$ is sampled from the suspicious region, $\mathbf H_i^{(\ell)}$ is generated according to the Rician channel model, and $P_j^{(\ell)}\sim U(0,P_{j,\max})$. 
For each sample, the missed detection probability, the auxiliary participation probability, and the localization failure probability are evaluated. The average LEP is then approximated as
\begin{equation}
    \widehat P_e(P_{j,\max},\mathbf W_j)=\frac{1}{L}\sum_{\ell=1}^{L}P_e^{(\ell)}(P_{j,\max},\mathbf W_j).
\end{equation}

Accordingly, $\mathbf P_1$ can be approximated by

\begin{subequations}
\begin{align}
        \widehat{\mathbf P}_1:\quad
\max_{P_{j,\max},\mathbf W_j}\quad
&\widehat P_e(P_{j,\max},\mathbf W_j)\\
\mathrm{s.t.}\quad
&\gamma_B(P_{j,\max},\mathbf W_j)\ge \gamma_{\mathrm{th}},\\
&P_s+P_{j,\max}\le P_A^{\max},\\
&P_{j,\max}\ge 0,\\
&\mathrm{tr}(\mathbf W_j\mathbf W_j^H)=1.
\end{align}
\end{subequations}

To facilitate the optimization of the jamming direction, we define the covariance matrix of jamming as $\mathbf Q_j=\mathbf W_j\mathbf W_j^H$. 
Then, $\mathbf Q_j\succeq \mathbf 0$ and $\mathrm{tr}(\mathbf Q_j)=1$.
The received jamming power at $\mathrm{Willie}_i$ can be rewritten as $\Vert\mathbf H_i\mathbf W_j\Vert_F^2=\mathrm{tr}\left(\mathbf H_i\mathbf Q_j\mathbf H_i^H\right)$. 
Similarly, the jamming leakage to Bob after receive combining becomes $\Vert\mathbf u_B^H\mathbf H_B\mathbf W_j\Vert^2=\mathrm{tr}\left(\mathbf H_B^H\mathbf u_B\mathbf u_B^H\mathbf H_B\mathbf Q_j\right)$.
Therefore, the optimization problem can be rewritten in terms of $P_{j,\max}$ and $\mathbf Q_j$ as
\begin{subequations}
\begin{align}
\mathbf P_2:\quad
\max_{P_{j,\max},\mathbf Q_j}\quad
&\widehat{P}_e(P_{j,\max},\mathbf Q_j)\\
\mathrm{s.t.}\quad
&\gamma_B(P_{j,\max},\mathbf Q_j)\ge \gamma_{\mathrm{th}},\\
&P_s+P_{j,\max}\le P_A^{\max},\\
&P_{j,\max}\ge 0,\\
&\mathbf Q_j\succeq \mathbf 0,\\
&\mathrm{tr}(\mathbf Q_j)=1.
\end{align}
\end{subequations}
If a low-rank jamming precoder is required, $\mathbf W_j$ can be recovered from $\mathbf Q_j$ by eigenvalue decomposition.

For a fixed jamming covariance matrix $\mathbf Q_j$, the maximum feasible $P_{j,\max}$ is determined by both Alice's maimum transmit power and Bob's minimum SINR requirement.
Define $G_B=|\mathbf u_B^H\mathbf H_B\mathbf w_s|^2$ and $J_B(\mathbf Q_j)=\mathrm{tr}\left(\mathbf H_B^H\mathbf u_B\mathbf u_B^H\mathbf H_B\mathbf Q_j\right)$.
Then Bob's SINR constraint becomes
\begin{equation}
    \frac{P_sG_B}{P_{j,\max}J_B(\mathbf Q_j)+\sigma_B^2}\ge\gamma_{\mathrm{th}}.
\end{equation}
If $J_B(\mathbf Q_j)>0$, this yields 
\begin{equation}
    P_{j,\max}\le\frac{P_sG_B/\gamma_{\mathrm{th}}-\sigma_B^2}{J_B(\mathbf Q_j)}.
\end{equation}
Therefore, the feasible maximum jamming power is
\begin{equation}
    P_{j,\max}^{\mathrm{fea}}(\mathbf Q_j)=\min\left\{P_A^{\max}-P_s,\frac{P_sG_B/\gamma_{\mathrm{th}}-\sigma_B^2}{J_B(\mathbf Q_j)}\right\}.
\end{equation}
If $J_B(\mathbf Q_j)=0$, the jamming signal lies in the effective null space of Bob, and the Bob SINR constraint does not limit $P_{j,\max}$. In this case,
\begin{equation}
    P_{j,\max}^{\mathrm{fea}}(\mathbf Q_j)=P_A^{\max}-P_s.
\end{equation}
This result suggests that Alice should steer the jamming signal toward the Willie satellites while suppressing its leakage to Bob.

Due to the non-convexity of $\mathbf P_2$, we adopt an alternating optimization framework. 
For a fixed $\mathbf Q_j$, $P_{j,\max}$ is optimized over the interval $0\le P_{j,\max}\le P_{j,\max}^{\mathrm{fea}}(\mathbf Q_j)$.
Since the objective $\widehat P_e(P_{j,\max},\mathbf Q_j)$ is one-dimensional with respect to $P_{j,\max}$, it can be optimized by a line search method, such as golden-section search or exhaustive grid search.

For a fixed $P_{j,\max}$, the covariance matrix $\mathbf Q_j$ of jamming is updated by solving
\begin{subequations}
\label{sub_P1}
    \begin{align}
\max_{\mathbf Q_j}\quad
&\widehat P_e(P_{j,\max},\mathbf Q_j)\\
\mathrm{s.t.}\quad
&\gamma_B(P_{j,\max},\mathbf Q_j)\ge \gamma_{\mathrm{th}},\\
&\mathbf Q_j\succeq \mathbf 0,\\
&\mathrm{tr}(\mathbf Q_j)=1.
\end{align}
\end{subequations}

The problem~\eqref{sub_P1} is still non-convex because $\widehat P_e$ contains nonlinear terms. It can be solved using projected gradient ascent. Specifically, $\mathbf Q_j$ is first updated along the ascent direction of $\widehat P_e$, and then projected onto the feasible set defined by the positive semidefinite constraint, the trace constraint, and Bob's SINR constraint.

\textbf{\textit{Remark:}}A low-complexity artificial jamming design under a special case can be obtained when Alice has sufficient spatial degrees of freedom to avoid interfering with Bob. Specifically, if the null space of the effective Bob channel is non-empty, i.e., $\mathrm{null}(\mathbf u_B^H\mathbf H_B)\neq \varnothing$, 
Alice can choose the jamming precoder such that $\mathbf u_B^H\mathbf H_B\mathbf W_j=\mathbf 0$.
In this case, the artificial jamming signal causes no interference to Bob after receive combining, and Bob's SINR reduces to $\gamma_B=\frac{P_s|\mathbf u_B^H\mathbf H_B\mathbf w_s|^2}{\sigma_B^2}$.
The optimal maximum transmit power of jamming is then only limited by Alice's peak power constraint and given by $P_{j,\max}^{*}=P_A^{\max}-P_s$.


It is also important to note that the impact of jamming on the missed detection probability depends on the adopted detector model. If the Willie performs energy detection on the total received emission, increasing $P_j$ may increase the received energy under $\mathcal{H}_1$, which can reduce the missed detection probability. 
In this case, jamming mainly improves location protection by increasing TDOA errors and degrading localization accuracy. 

\section{Numerical Results}

This section presents extensive simulation results to illustrate the impact of key parameters on the performance of the proposed method and provide verification of the theoretical analysis. Unless otherwise specified, we set the values of other parameters are as follows. Alice is equipped with $N_A=6$ antennas, Bob is equipped with $N_B=4$ antennas, and each Willie is equipped with $N_{W_i}=4$ antennas. The false alarm probability is $P_{fa}=0.05$. The noise power is normalized as $\sigma_W^2=1$. The suspicious region radius is $R_A=50$ km, the satellite altitude is $300$ km. The noise power at Bob is normalized as $\sigma_B^2=1$. The Rician factor of the Alice-Bob link is set to $K_B=6$ dB. We perform $20000$ Monte Carlo trials.

\begin{figure}
	\centering
	\includegraphics[width=3.5in]{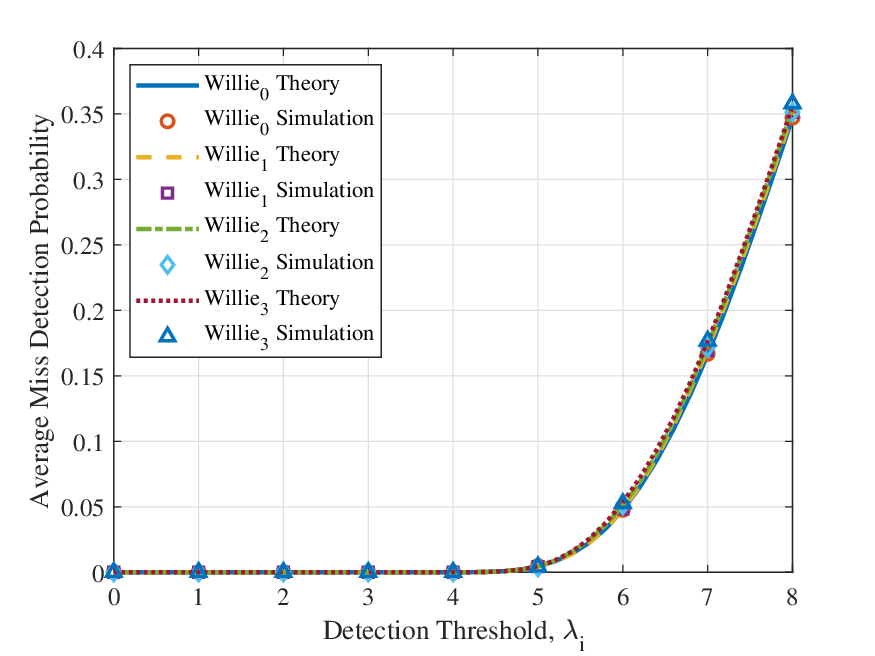}
	\caption{The impact of detection threshold on the average missed detection probability of each warden satellite.}
	\label{Fig.2}
\end{figure}

To investigate the impact of the detection threshold on the average missed detection probability, we summarize in Fig.~\ref{Fig.2} to illustrate how $\bar{P}_{md,i}$ varies with $\lambda_{i}$ for the following settings. We consider a adversarial satellites system with four Willies, i.e., $\mathrm{Willie}_0,\mathrm{Willie}_1,\mathrm{Willie}_2$, and $\mathrm{Willie}_3$. The maximum jamming power is $P_{j,\max}=1$ W and the transmit power of information signal is $P_s=1$ W,. 
The Rician factors of the four Willie links are set to $3$, $2$, $1$, and $0$ dB, respectively.

 It can be observed from Fig.~\ref{Fig.2} that the theoretical results match well with the simulation results for all four Willies, indicating the correctness of our theoretical analysis for the average missed detection probability.
 We can also see that the average missed detection probability increases with the detection threshold. This is because a larger threshold makes the energy detector easier in declaring $\mathcal{H}_0$, thereby increasing the missed detection probability. When the detection threshold is small, all Willies can easily declare $\mathcal{H}_1$, and the corresponding missed detection probability is close to zero. As the threshold increases, the received energy under $\mathcal{H}_1$ is more likely to fall below the threshold, resulting in a rapid increase in the missed detection probability.

\begin{figure}
	\centering
	\includegraphics[width=3.5in]{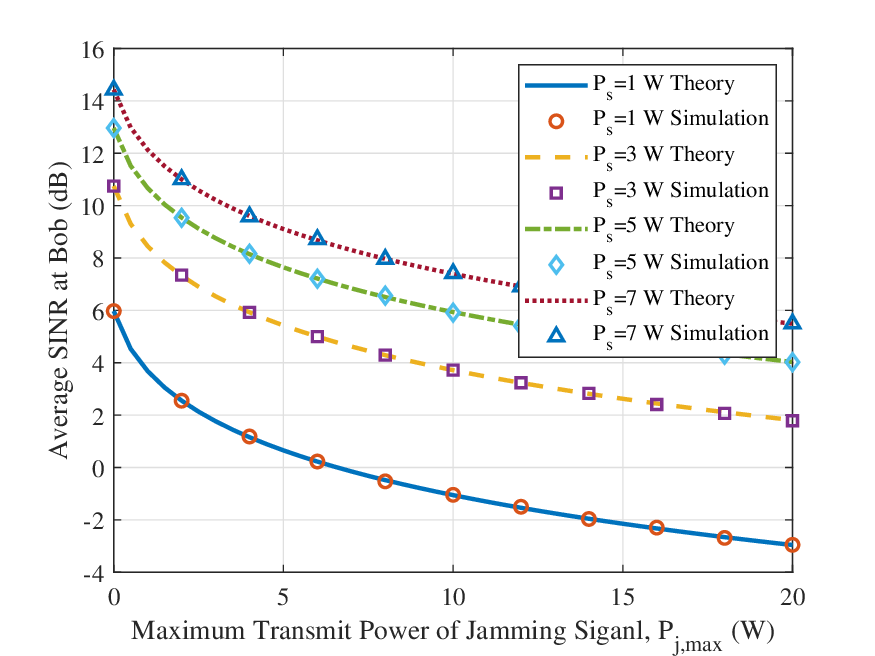}
	\caption{The impact of $P_{j}^{\max}$ on the average SINR at Bob.}
	\label{Fig.3}
\end{figure}

To investigate the impact of the maximum transmit power of jamming signal $P_{j,\max}$ on the legitimate Alice-Bob link, we evaluate the average SINR at Bob under different transmit powers of Alice's information signal for the transmit power of the information signal is set to $P_s=\{1, 3, 5, 7 \}$W. The results are presented in Fig.~\ref{Fig.3}. 
It can be observed from Fig.~\ref{Fig.3} that, for all values of $P_s$, the average SINR at Bob decreases as $P_{j,\max}$ increases. This is because a larger $P_{j,\max}$ increases the average interference leakage from Alice's jamming signal to Bob, thereby enlarging the denominator of Bob's SINR in~\eqref{SINR_B}. 
For a given $P_{j,\max}$, the average SINR at Bob increases with the transmit power of Alice's information signal $P_s$. This is because a larger $P_s$ directly improves the useful signal power received at Bob, while the jamming leakage and noise power remain unchanged for a fixed $P_{j,\max}$. Therefore, increasing $P_s$ can increase the numerator of the  Bob's SINR in~\eqref{SINR_B}.
In addition, the theoretical results match closely with the simulation results for all considered values of $P_s$, which validates the correctness of the theoretical SINR analysis. These results reveal an inherent tradeoff in the location protection wireless communication network. Increasing $P_{j,\max}$ can enhance Alice's ability to interfere with the Willie system, but it may also degrade the legitimate communication quality at Bob. Therefore, the jamming power should be carefully designed under Bob's SINR constraint.

\begin{figure}
	\centering
	\includegraphics[width=3.5in]{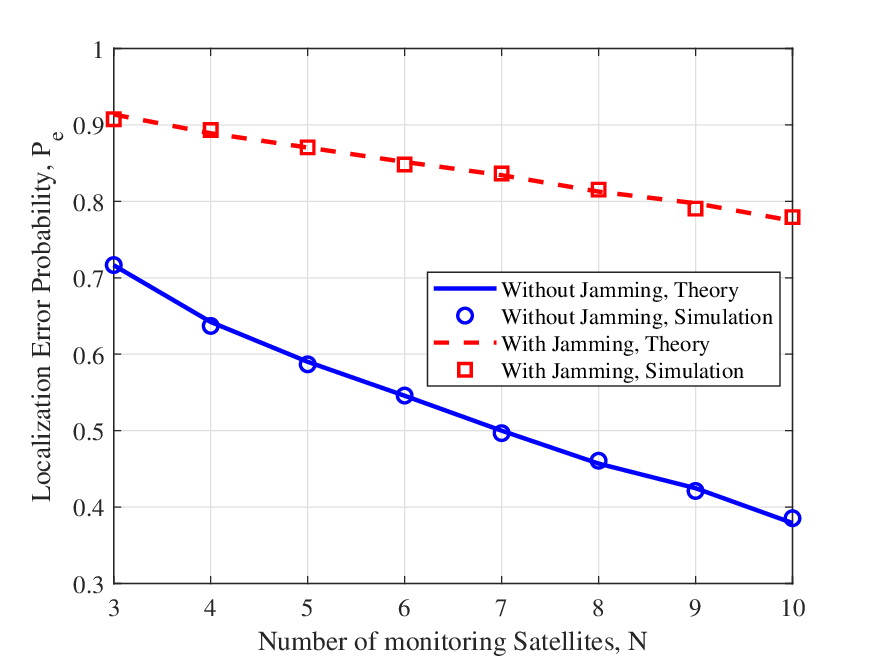}
	\caption{The impact of the number of warden satellites on the localization error probability.}
	\label{Fig.5}
\end{figure}

To investigate the impact of the number of adversarial satellites on the LEP, we evaluate how $P_e$ varies with the total number of Willies $N$, as well as we compare the cases with the cooperative jamming and without jamming. 
The results are presented in Fig.~\ref{Fig.5} for the setting of key parameters as follows.
The transmit power of Alice's information signal is $P_s=3$ W and the localization error threshold is set to $d_{th}=800$ m.
 For the jamming case, the maximum jamming power is set to $P_{j,\max}=10$ W, while $P_{j,\max}=0$ W is adopted for the without jamming case.
 
 We can be observe from Fig.~\ref{Fig.5} that $P_e$ decreases as $N$ increases for both the with jamming and without jamming cases. This is because a larger number of adversarial satellites provides more auxiliary Willies that may participate in TDOA localization. As a result, the probability of having insufficient TDOA measurements decreases, and the localization geometry becomes more favorable. Therefore, the Willie system can localize Alice more accurately when more adversarial satellites are available, leading to a lower LEP.
Moreover, the LEP with jamming is consistently higher than that without jamming over the entire range of $N$. This demonstrates that the cooperative jamming effectively degrades the cooperative localization capability of the Willie system. Specifically, jamming increases the TDOA estimation error and enlarges the localization error, thereby increasing the probability that the localization error exceeds the threshold $d_{th}$. In addition, jamming may also reduce the effective participation of auxiliary Willies by degrading their detection performance. These two effects jointly improve Alice's location protection.
It is also observed that the gap between the with jamming and without jamming curves remains significant as $N$ increases. Although adding more adversarial satellites improves the Willie system's localization capability, the proposed jamming strategy still maintains a clear location protection gain. This indicates that random jamming can effectively counteract the benefit of increasing the number of cooperative adversarial satellites. Finally, the theoretical results closely match the simulation results, which validates the correctness of the derived localization error probability analysis.

\begin{figure}
	\centering
	\includegraphics[width=3.5in]{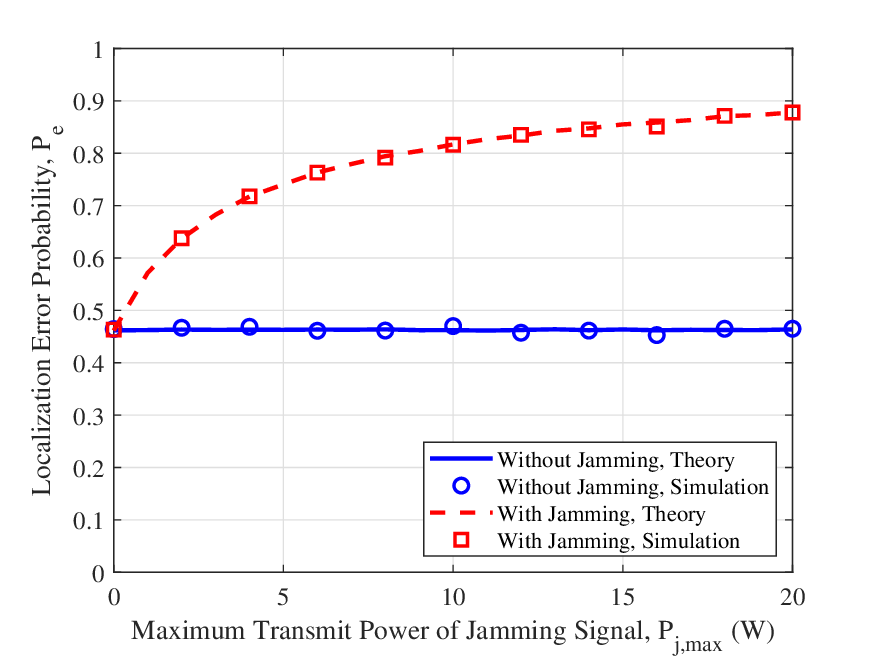}
	\caption{The impact of $P_{j}^{\max}$ on the localization error probability.}
	\label{Fig.6}
\end{figure}

To investigate the impact of the maximum transmit power of the jamming signal on the localization error probability, we summarize in Fig.~\ref{Fig.6} how $P_e$ varies with $P_{j,\max}$ under the with jamming and without jamming cases. The key parameters of the system are set as follows. The total number of adversarial satellites is set to $N=8$, the transmit power of Alice's information signal is set to $P_s=3$ W, and the localization error threshold is $d_{th}=800$ m.

 It can be observed from Fig.~\ref{Fig.6} that, in the without jamming case, $P_e$ remains almost unchanged as $P_{j,\max}$ increases. This is because no artificial jamming is transmitted in this baseline case, and thus the Willie system's detection and TDOA localization performance are independent of $P_{j,\max}$. 
 In contrast, in the with jamming case, $P_e$ increases significantly with $P_{j,\max}$. When $P_{j,\max}$ is relatively small, the jamming signal only slightly degrades the TDOA measurement, and the LEP is close to that of the no jamming baseline. As $P_{j,\max}$ continue to increase, the jamming signal introduces stronger interference at the adversarial satellites, which reduces the effective SINR for TDOA estimation and enlarges the localization error. Consequently, the probability that the localization error exceeds the threshold $d_{th}$ increases.
Nevertheless, the results under the with jamming case remains consistently higher than that under the without jamming case over the entire range of $P_{j,\max}$, which confirms the effectiveness of the proposed jamming strategy in degrading the cooperative TDOA localization capability of the Willie system.

\begin{figure}
	\centering
	\includegraphics[width=3.5in]{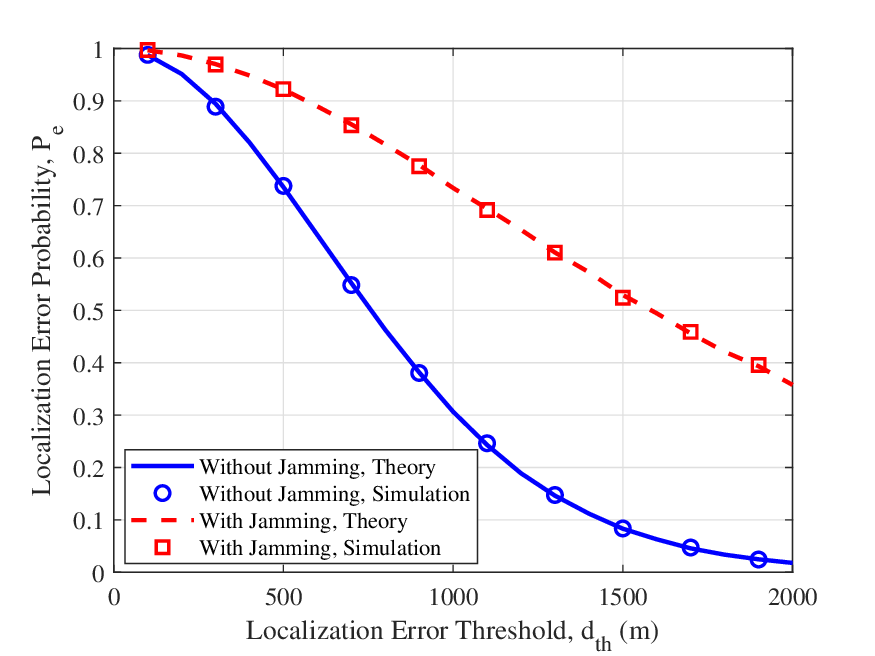}
	\caption{The impact of the localization error threshold $d_{\mathrm{th}}$ on the localization error probability.}
	\label{Fig.7}
\end{figure}

To investigate the impact of the localization error threshold $d_{th}$ on the localization error probability, we evaluate $P_e$ under different values of $d_{th}$ and compare the cases with and without jamming. The results are illustrated in Fig.~\ref{Fig.7} to show how the LEP $P_e$ varies with the localization error threshold $d_{th}$ for the follow settings.
The total number of adversarial satellites is set to $N=8$. The transmit power of Alice's information signal is set to $P_s=3$ W.
For the with jamming case, the maximum jamming power is set to $P_{j,\max}=10$ W, while $P_{j,\max}=0$ W is adopted for the without jamming case. 

 It can be observed from Fig.~\ref{Fig.7} that $P_e$ decreases as $d_{th}$ increases for both the with jamming and without jamming cases. This is because a larger $d_{th}$ relaxes the criterion for declaring localization failure. In other words, when $d_{th}$ is relatively small, the Willie system is hard to satisfy the location error threshold, even a small localization error is regarded as a failure, resulting in a high LEP. As $d_{th}$ increases, the Willie system is more likely to satisfy the required localization accuracy, and thus $P_e$ decreases.
Moreover, the results under the jamming case is consistently higher than that under the without jamming case over the entire range of $d_{th}$. 
This demonstrates that the cooperative jamming signal effectively degrades the TDOA-based localization capability of the Willie system. 
Specifically, random jamming reduces the effective SINR for TDOA estimation, increases the TDOA error, and enlarges the localization error. As a result, for the same localization threshold, the probability that the estimated position deviates from Alice's true location by more than $d_{th}$ becomes significantly higher.
It is also worth noting that the performance gap between the two curves is more pronounced when $d_{th}$ is relatively large. Without jamming, the localization error probability rapidly decreases as the threshold increases, indicating that the Willie system can achieve accurate localization when the allowable error range is relaxed. In contrast, with jamming, $P_e$ decreases much more slowly, which implies that the proposed jamming strategy can maintain a high level of location protection even under a relatively large localization error threshold. 

\begin{figure}
	\centering
	\includegraphics[width=3.5in]{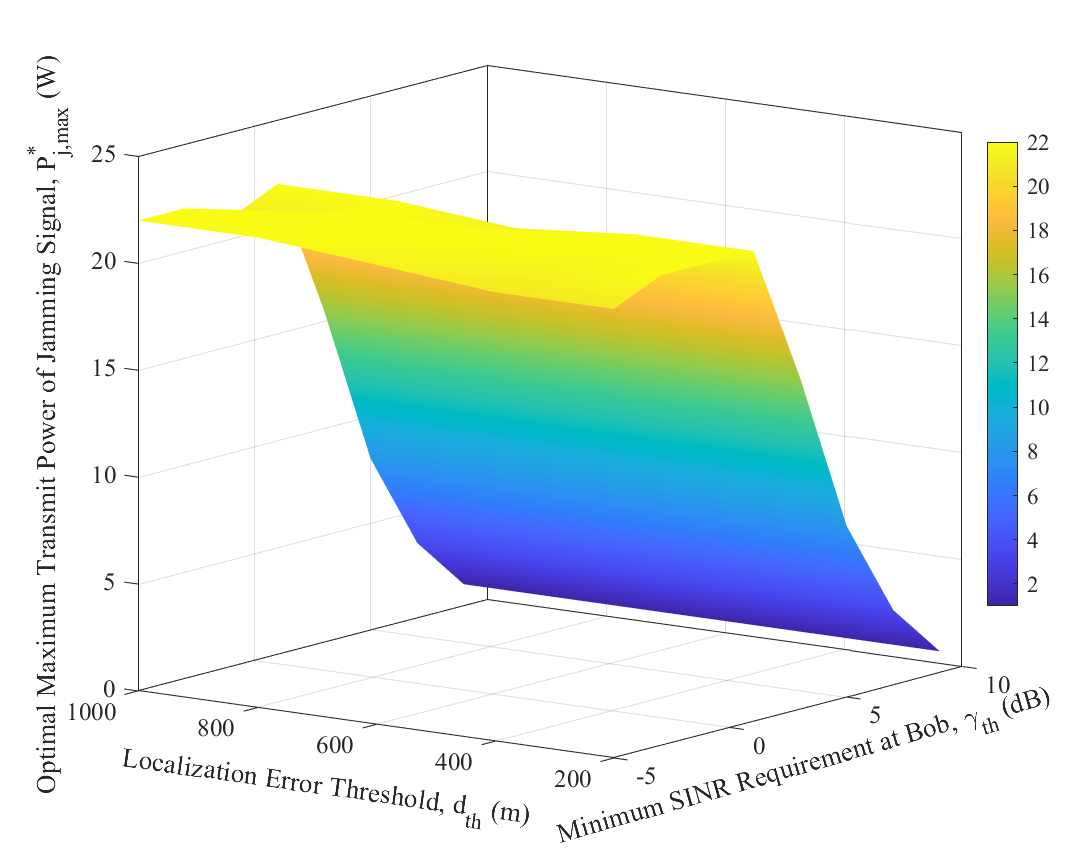}
	\caption{The relationship between location protection and reliable communication.}
	\label{Fig.8}
\end{figure}

To investigate the joint impact of the localization error threshold $d_{th}$ and Bob's minimum SINR requirement $\gamma_{th}$ on the optimal jamming power, we evaluate how the optimal maximum transmit power of the jamming signal $P_{j,\max}^{*}$ varies with $d_{th}$ and $\gamma_{th}$, which is summarized in Fig.~\ref{Fig.8}. 
In this experiment, 
The information signal power is set to $P_s=3$ W, the maximum transmit power at Alice is $P_A^{\max}=25$ W.
It can be observed from Fig.~\ref{Fig.8} that $P_{j,\max}^{*}$ decreases as $\gamma_{th}$ increases. This is because a larger $\gamma_{th}$ imposes a stricter communication quality requirement on the legitimate Alice-Bob link. To satisfy this constraint, Alice must reduce the maximum allowable jamming power so that the interference leakage to Bob does not excessively degrade the received SINR. When $\gamma_{th}$ is relatively small, the SINR requirement at Bob is loose, and the optimal maximum transmit power of the jamming signal is mainly limited by Alice's total transmit power $P_{A}^{\max}$. Therefore, $P_{j,\max}^{*}$ approaches its maximum feasible value determined by $P_A^{\max}-P_s$. In contrast, when $\gamma_{th}$ becomes large, the feasible maximum transmit power of the jamming signal is dominated by the Bob's minimum SINR requirement, resulting in a sharp reduction of $P_{j,\max}^{*}$.

It can also be observed from Fig.~\ref{Fig.8} that $P_{j,\max}^{*}$ is nearly unchanged along the $d_{th}$ dimension. This is because $d_{th}$ affects the value of the LEP $P_e$, but it does not directly affect the feasibility of the jamming power under the Bob's SINR and Alice's transmit power constraints. 
Since $P_e$ increases monotonically with $P_{j,\max}$, the optimal maximum transmit power of the jamming signal is to use the largest feasible maximum jamming power $P_{j,\max}^{\mathrm{fea}}$. As a result, the optimal $P_{j,\max}^{*}$ is primarily determined by $\gamma_{th}$ and the Alice's transmit power constraint, rather than by $d_{th}$. Nevertheless, a larger $d_{th}$ would require stronger jamming to achieve the same LEP level, which will be reflected in the achieved $P_e$ rather than in the optimal maximum transmit power $P_{j,\max}^{*}$ of the jamming signal.
Overall, this result reveals the fundamental tradeoff between location protection and legitimate communication reliability. A smaller $\gamma_{th}$ allows Alice to allocate more power to jamming, thereby providing more flexibility for location protection. Conversely, a stringent Bob SINR requirement significantly restricts the feasible jamming power and limits Alice’s ability to degrade the Willie system's TDOA localization performance.

\section{Conclusion}
This paper represents an initial attempt on the anti-localization uplink communication in a satellite-terrestrial system against cooperative TDOA-based localization. A cooperative jamming-based scheme for such anti-localization communication was proposed, and a new localization error probability (LEP) metric was also developed to fully depict the localization performance at adversarial satellites. Some interesting findings of this paper are as follows. Firstly, through a careful joint design of coding/power allocation at transmitter and combining vector-based signal processing at receiver, it is possible to apply cooperative jamming to achieve the reliable uplink communications in satellite-terrestrial systems while resisting the localization of transmitter from adversary satellites. Secondly, with the cooperative jamming-based anti-localization, the localization performance in terms of LEP at adversaries is sensitive to both the maximum jamming power and jamming precoding matrix, so a careful jamming design is crucial for forcing LEP to a high level. Thirdly, since a stringent requirement on the minimum SINR at Bob will unavoidably restrict the feasible jamming power and thus the achievable LEP, so a tradeoff should be initiated between the location protection guarantee and communication reliability.

\bibliographystyle{IEEEtran}
\bibliography{reference}

@article{al2022survey,
  title={A survey on nongeostationary satellite systems: The communication perspective},
  author={Al-Hraishawi, Hayder and Chougrani, Houcine and Kisseleff, Steven and Lagunas, Eva and Chatzinotas, Symeon},
  journal={IEEE Communications Surveys \& Tutorials},
  volume={25},
  number={1},
  pages={101--132},
  year={2022},
  publisher={IEEE}
}

@book{maral2020satellite,
  title={Satellite communications systems: systems, techniques and technology},
  author={Maral, G{\'e}rard and Bousquet, Michel and Sun, Zhili},
  year={2020},
  publisher={John Wiley \& Sons}
}

@article{li2024fundamentals,
  title={Fundamentals of satellite-maritime communications: Downlink and uplink analysis},
  author={Li, Zhuhang and Shang, Bodong},
  journal={IEEE Transactions on Communications},
  volume={73},
  number={4},
  pages={2191--2206},
  year={2024},
  publisher={IEEE}
}

@article{heo2023mimo,
  title={MIMO satellite communication systems: A survey from the PHY layer perspective},
  author={Heo, Jehyun and Sung, Seungwoo and Lee, Hyunwoo and Hwang, Incheol and Hong, Daesik},
  journal={IEEE Communications Surveys \& Tutorials},
  volume={25},
  number={3},
  pages={1543--1570},
  year={2023},
  publisher={IEEE}
}

@article{ho1993solution,
  title={Solution and performance analysis of geolocation by TDOA},
  author={Ho, KC and Chan, YT},
  journal={IEEE Transactions on Aerospace and Electronic Systems},
  volume={29},
  number={4},
  pages={1311--1322},
  year={1993},
  publisher={IEEE}
}

@inproceedings{kaune2012accuracy,
  title={Accuracy studies for TDOA and TOA localization},
  author={Kaune, Regina},
  booktitle={2012 15th International Conference on Information Fusion},
  pages={408--415},
  year={2012},
  organization={IEEE}
}

@article{liang2012tdoa,
  title={TDoA for passive localization: Underwater versus terrestrial environment},
  author={Liang, Qilian and Zhang, Baoju and Zhao, Chenglin and Pi, Yiming},
  journal={IEEE Transactions on Parallel and Distributed Systems},
  volume={24},
  number={10},
  pages={2100--2108},
  year={2012},
  publisher={IEEE}
}

@INPROCEEDINGS{5977569,
  author={Kaune, Regina and Hörst, Julian and Koch, Wolfgang},
  booktitle={14th International Conference on Information Fusion}, 
  title={Accuracy analysis for TDOA localization in sensor networks}, 
  year={2011},
  volume={},
  number={},
  pages={1-8},
  doi={}}

@article{lin2020secure,
  title={Secure and energy efficient transmission for RSMA-based cognitive satellite-terrestrial networks},
  author={Lin, Zhi and Lin, Min and Champagne, Benoit and Zhu, Wei-Ping and Al-Dhahir, Naofal},
  journal={IEEE Wireless Communications Letters},
  volume={10},
  number={2},
  pages={251--255},
  year={2020},
  publisher={IEEE}
}

@article{huang2023deep,
  title={Deep reinforcement learning-based resource allocation for RSMA in LEO satellite-terrestrial networks},
  author={Huang, Jingfei and Yang, Yang and Lee, Jemin and He, Dazhong and Li, Yonghui},
  journal={IEEE Transactions on Communications},
  volume={72},
  number={3},
  pages={1341--1354},
  year={2023},
  publisher={IEEE}
}

@article{castro2007cross,
  title={Cross-layer packet scheduler design of a multibeam broadband satellite system with adaptive coding and modulation},
  author={Castro, M Angeles V{\'a}zquez and Granados, Gonzalo Seco},
  journal={IEEE Transactions on Wireless Communications},
  volume={6},
  number={1},
  pages={248--258},
  year={2007},
  publisher={IEEE}
}

@article{you2020massive,
  title={Massive MIMO transmission for LEO satellite communications},
  author={You, Li and Li, Ke-Xin and Wang, Jiaheng and Gao, Xiqi and Xia, Xiang-Gen and Ottersten, Bj{\"o}rn},
  journal={IEEE Journal on selected areas in communications},
  volume={38},
  number={8},
  pages={1851--1865},
  year={2020},
  publisher={IEEE}
}

@inproceedings{bodenhausen2023securing,
  title={Securing wireless communication in critical infrastructure: Challenges and opportunities},
  author={Bodenhausen, J{\"o}rn and Sorgatz, Christian and Vogt, Thomas and Grafflage, Kolja and R{\"o}tzel, Sebastian and Rademacher, Michael and Henze, Martin},
  booktitle={International Conference on Mobile and Ubiquitous Systems: Computing, Networking, and Services},
  pages={333--352},
  year={2023},
  organization={Springer}
}

@article{lorincz2004sensor,
  title={Sensor networks for emergency response: challenges and opportunities},
  author={Lorincz, Konrad and Malan, David J and Fulford-Jones, Thaddeus RF and Nawoj, Alan and Clavel, Antony and Shnayder, Victor and Mainland, Geoffrey and Welsh, Matt and Moulton, Steve},
  journal={IEEE pervasive Computing},
  volume={3},
  number={4},
  pages={16--23},
  year={2004},
  publisher={IEEE}
}

@article{torrieri2007statistical,
  title={Statistical theory of passive location systems},
  author={Torrieri, Don J},
  journal={IEEE transactions on Aerospace and Electronic Systems},
  number={2},
  pages={183--198},
  year={2007},
  publisher={IEEE}
}

@article{hao2020interference,
  title={Interference geolocation in satellite communications systems: An overview},
  author={Hao, Caiyong and Feng, Daquan and Zhang, Qinyu and Xia, Xiang-Gen},
  journal={IEEE Vehicular Technology Magazine},
  volume={16},
  number={1},
  pages={66--74},
  year={2020},
  publisher={IEEE}
}

@inproceedings{clements2023dual,
  title={Dual-satellite geolocation of terrestrial GNSS jammers from low Earth orbit},
  author={Clements, Zachary and Humphreys, Todd E and Ellis, Patrick},
  booktitle={2023 IEEE/ION Position, Location and Navigation Symposium (PLANS)},
  pages={458--469},
  year={2023},
  organization={IEEE}
}

@article{chan1994simple,
  title={A simple and efficient estimator for hyperbolic location},
  author={Chan, Yiu Tong and Ho, Kenneth C},
  journal={IEEE transactions on signal processing},
  volume={42},
  number={8},
  pages={1905--1915},
  year={1994},
  publisher={IEEE}
}

@article{JNaNA1,
    author ={Wang, Sibo and Wang, Wu},
    title = {A Review of Satellite Passive Localization: Principles, Parameter Estimation, and Positioning Algorithms},
    journal = {Journal of Networking and Network Applications},
    volume={5},
    number={3},
    pages={110-119},
    year = {2025},
}

@inproceedings{kaune2011accuracy,
  title={Accuracy analysis for TDOA localization in sensor networks},
  author={Kaune, Regina and H{\"o}rst, Julian and Koch, Wolfgang},
  booktitle={14th international conference on information fusion},
  pages={1--8},
  year={2011},
  organization={IEEE}
}

@ARTICLE{8964427,
  author={Zou, Yanbin and Liu, Huaping},
  journal={IEEE Communications Letters}, 
  title={TDOA Localization With Unknown Signal Propagation Speed and Sensor Position Errors}, 
  year={2020},
  volume={24},
  number={5},
  pages={1024-1027},
  doi={10.1109/LCOMM.2020.2968434}}

@article{urkowitz1967energy,
  title={Energy detection of unknown deterministic signals},
  author={Urkowitz, Harry},
  journal={Proceedings of the IEEE},
  volume={55},
  number={4},
  pages={523--531},
  year={1967},
  publisher={IEEE}
}

@article{bodenham2016comparison,
  title={A comparison of efficient approximations for a weighted sum of chi-squared random variables},
  author={Bodenham, Dean A and Adams, Niall M},
  journal={Statistics and Computing},
  volume={26},
  number={4},
  pages={917--928},
  year={2016},
  publisher={Springer}
}

@article{satterthwaite1946approximate,
  title={An approximate distribution of estimates of variance components},
  author={Satterthwaite, Franklin E},
  journal={Biometrics bulletin},
  volume={2},
  number={6},
  pages={110--114},
  year={1946},
  publisher={JSTOR}
}

@article{atapattu2011energy,
  title={Energy detection based cooperative spectrum sensing in cognitive radio networks},
  author={Atapattu, Saman and Tellambura, Chintha and Jiang, Hai},
  journal={IEEE Transactions on wireless communications},
  volume={10},
  number={4},
  pages={1232--1241},
  year={2011},
  publisher={IEEE}
}

@article{li2011quantitative,
  title={Quantitative measurement and design of source-location privacy schemes for wireless sensor networks},
  author={Li, Yun and Ren, Jian and Wu, Jie},
  journal={IEEE Transactions on Parallel and Distributed Systems},
  volume={23},
  number={7},
  pages={1302--1311},
  year={2011},
  publisher={IEEE}
}

@article{mahmoud2011cloud,
  title={A cloud-based scheme for protecting source-location privacy against hotspot-locating attack in wireless sensor networks},
  author={Mahmoud, Mohamed MEA and Shen, Xuemin},
  journal={IEEE Transactions on Parallel and Distributed Systems},
  volume={23},
  number={10},
  pages={1805--1818},
  year={2011},
  publisher={IEEE}
}

@article{wang2019source,
  title={Source-location privacy protection based on anonymity cloud in wireless sensor networks},
  author={Wang, Na and Fu, Junsong and Li, Jian and Bhargava, Bharat K},
  journal={IEEE Transactions on Information Forensics and Security},
  volume={15},
  pages={100--114},
  year={2019},
  publisher={IEEE}
}

@article{han2019cpslp,
  title={CPSLP: A cloud-based scheme for protecting source location privacy in wireless sensor networks using multi-sinks},
  author={Han, Guangjie and Miao, Xu and Wang, Hao and Guizani, Mohsen and Zhang, Wenbo},
  journal={IEEE Transactions on Vehicular Technology},
  volume={68},
  number={3},
  pages={2739--2750},
  year={2019},
  publisher={IEEE}
}

@article{wang2019probabilistic,
  title={A probabilistic source location privacy protection scheme in wireless sensor networks},
  author={Wang, Hao and Han, Guangjie and Zhang, Wenbo and Guizani, Mohsen and Chan, Sammy},
  journal={IEEE Transactions on Vehicular Technology},
  volume={68},
  number={6},
  pages={5917--5927},
  year={2019},
  publisher={IEEE}
}

@article{han2019dynamic,
  title={A dynamic multipath scheme for protecting source-location privacy using multiple sinks in WSNs intended for IIoT},
  author={Han, Guangjie and Wang, Hao and Miao, Xu and Liu, Li and Jiang, Jinfang and Peng, Yan},
  journal={IEEE Transactions on Industrial Informatics},
  volume={16},
  number={8},
  pages={5527--5538},
  year={2019},
  publisher={IEEE}
}

@article{win2022location,
  title={Location awareness via intelligent surfaces: A path toward holographic NLN},
  author={Win, Moe Z and Wang, Ziyi and Liu, Zhenyu and Shen, Yuan and Conti, Andrea},
  journal={IEEE Vehicular Technology Magazine},
  volume={17},
  number={2},
  pages={37--45},
  year={2022},
  publisher={IEEE}
}

@article{wymeersch2020radio,
  title={Radio localization and mapping with reconfigurable intelligent surfaces: Challenges, opportunities, and research directions},
  author={Wymeersch, Henk and He, Jiguang and Denis, Benoit and Clemente, Antonio and Juntti, Markku},
  journal={IEEE Vehicular Technology Magazine},
  volume={15},
  number={4},
  pages={52--61},
  year={2020},
  publisher={IEEE}
}

@article{emenonye2023ris,
  title={RIS-aided localization under position and orientation offsets in the near and far field},
  author={Emenonye, Don-Roberts and Dhillon, Harpreet S and Buehrer, R Michael},
  journal={IEEE Transactions on Wireless Communications},
  volume={22},
  number={12},
  pages={9327--9345},
  year={2023},
  publisher={IEEE}
}


 





\end{document}